\documentclass[12pt,a4paper,final]{iopart}

\usepackage{iopams}  
\usepackage{graphicx}
\usepackage[breaklinks=true,colorlinks=true,linkcolor=blue,urlcolor=blue,citecolor=blue]{hyperref}

\usepackage{bm}
\usepackage[normalem]{ulem}
\usepackage{color,hyperref}

\begin{document}

\title{Two-Dimensionally Confined Topological Edge States in Photonic Crystals}

\author{Sabyasachi Barik$^{1,2}$, Hirokazu Miyake$^{2}$, Wade DeGottardi$^{2}$, Edo Waks$^{2,3}$, Mohammad Hafezi$^{2,3,4}$}

\address{$^1$Department of Physics, University of Maryland, College Park, MD 20742, USA}
\address{$^2$Joint Quantum Institute, University of Maryland, College Park, MD 20742, USA and National Institute of Standards and Technology, Gaithersburg, MD 20899, USA}
\address{$^3$Department of Electrical and Computer Engineering and Institute for Research in Electronics and Applied Physics, University of Maryland, College Park, Maryland 20742, USA}
\address{$^4$Kavli Institute of Theoretical Physics, Santa Barbara, CA 93106, USA}

\ead{sbarik@umd.edu}
\vspace{10pt}

\begin{abstract}
We present an all-dielectric photonic crystal structure that supports two-dimensionally confined helical topological edge states. The topological properties of the system are controlled by the crystal parameters. An interface between two regions of differing band topologies gives rise to topological edge states confined in a dielectric slab that propagate around sharp corners without backscattering. Three dimensional finite-difference time-domain calculations show these edges to be confined in the out-of-plane direction by total internal reflection. Such nanoscale photonic crystal architectures could enable strong interactions between photonic edge states and quantum emitters. 
\end{abstract}


\maketitle

\section{Introduction}

Topology is a ubiquitous concept in physics, ranging from electrons in solid state~\cite{hasan10,qi11}, quantum degenerate gases~\cite{galitski13,jotzu14}, and sound~\cite{yang15,peano15,wang15,huber15}. A key manifestation of topological physics is the presence of edge modes which are robust to local disorder. The prospect of using topological photonic materials for such robust propagation of light has attracted a great deal of interest~\cite{lu14,hafezi14}.

Topologically-protected edge states have been experimentally demonstrated in systems at microwave freqencies~\cite{wang09,khanikev16} and optical frequencies, specifically in ring resonators~\cite{hafezi13, mittal14}, and in coupled waveguides ~\cite{rechtsman13}. Subsequent work measured the invariants characterizing the topology of two-dimensional photonic systems~\cite{mittal16}. Embedding quantum emitters into these optical frequency devices could generate strong optical non-linearities that exhibit new physical behavior. Theoretical work has shown that the interplay between emitters and chiral states results in intriguing phenomena such as many-body position-independent scattering~\cite{gritsev14}, dimerization of driven emitters~\cite{pichler15} and fractional quantum Hall states~\cite{cho08,carusotto12,hafezi13b}. 

Strong light-matter interactions with optical emitters usually require the concentration of light to small mode-volume nanophotonic devices~\cite{joannopoulos08}. Two-dimensional photonic crystals are one of the most promising nanophotonic platforms for this application because they confine light to less than an optical wavelength~\cite{polman15,lodahl15}. Recently, several works have proposed photonic crystal structures where deformations open a gap in the Dirac cone dispersion to achieve non-trivial topological bands~\cite{wu15,ma15,ma16,longphotonic,dong15}. However, these proposals either make use of dielectric cylinders, which make it difficult to experimentally achieve out-of-plane confinement in planar all-dielectric nanophotonic systems, or make use of metallic mirrors to achieve out-of-plane confinement, which is undesirable for devices operating at optical frequencies where loss in metals is significant. Thus, it would be highly desirable to create an all-dielectric topological photonic crystal which is confined in the out-of-plane direction without the use of metal.

\begin{figure}[htbp]
\centering
\includegraphics[width=0.8\columnwidth]{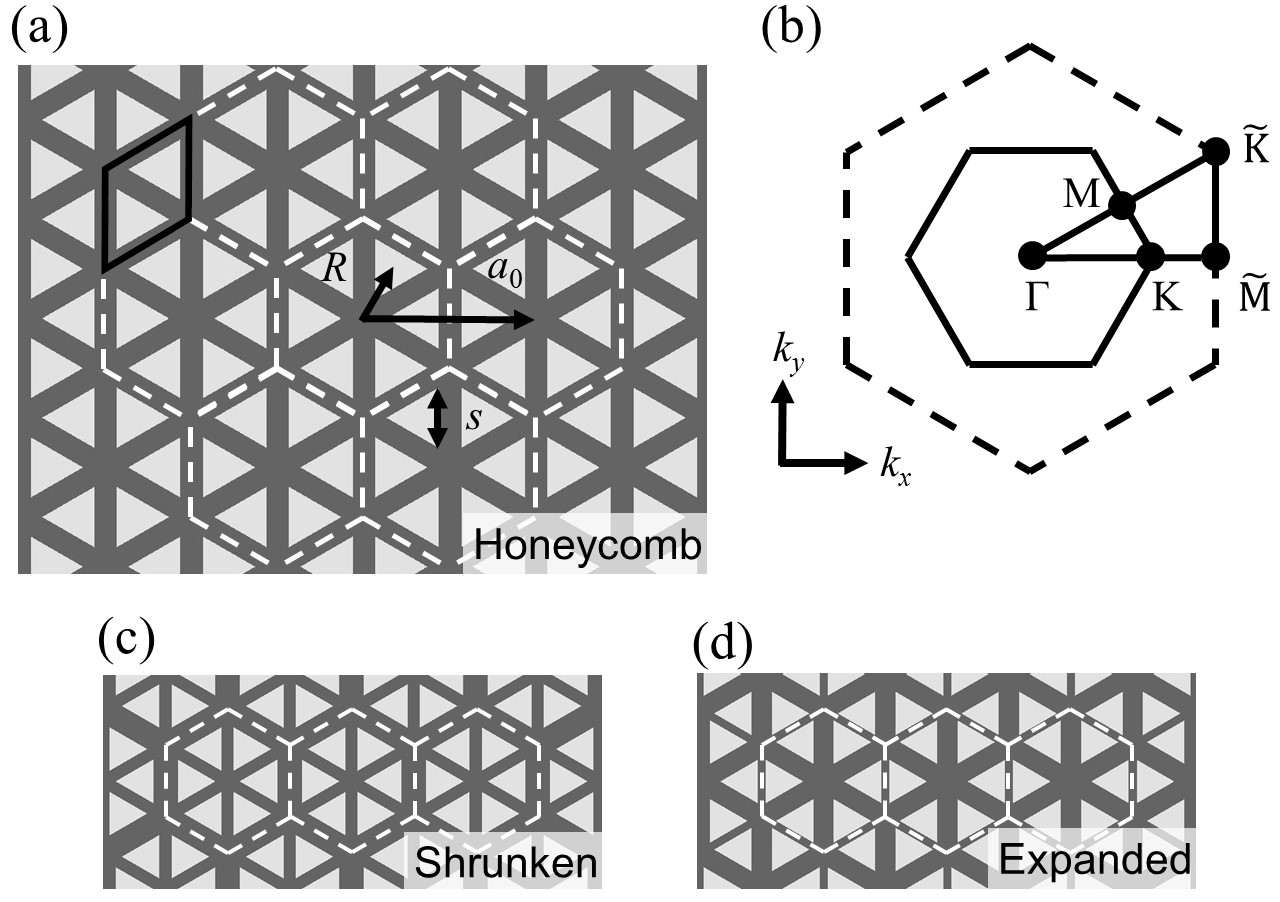}
\caption{\label{fig:structure} Schematic of our proposed honeycomb-lattice-like photonic crystal. (a) Baseline structure of equilateral triangular holes arranged in a honeycomb lattice in a dielectric material. This honeycomb lattice can be viewed as a triangular lattice of two-hole unit cells (black solid rhombus), or alternatively as a triangular lattice of six-hole unit cells (white dashed hexagons), which we call honeycomb clusters with $R=a_0/3$. (b) First Brillouin zones for the six-hole (solid) and two-hole (dashed) unit cells. The letters indicate high-symmetry points.  (c) [and (d)] Same structure as in (a) except that $R<a_0/3$ ($R>a_0/3)$, which we call shrunken (expanded) clusters.}

\end{figure}

Here we demonstrate that an all-dielectric topological photonic crystal can exhibit two-dimensional edge states confined by total internal reflection in a dielectric slab, enabling low-loss confinement of light in the third dimension. This structure addresses the challenge of experimentally realizing topological photonic crystals and enabling strong interactions with optical emitters. Our system exhibits spin quantum Hall physics for pseudo-spin photonic polarizations. As a result of time-reversal symmetry, the edge states are helical: edge states of opposite helicity travel in opposite directions. We utilize a honeycomb periodic structure with six-fold symmetry based on triangular holes.  This structure ensures a complete bandgap for transverse-electric-like modes. Deformations of the unit cell that preserve its rotational symmetry change the topology of the structure. We show that interfacing two materials of different band topologies results in robust two-dimensionally confined edge states that can propagate around sharp bends.  

\section{Photonic Crystal Design and Band Structure}

Fig.~\ref{fig:structure} shows a schematic of our photonic crystal structure. The starting point is a honeycomb lattice made of equilateral triangular holes in a dielectric material as shown in Fig.~\ref{fig:structure}(a). We can view this system as a triangular lattice with a basis consisting of two triangular holes, as is typically done in studies of graphene~\cite{neto09}. The black outline shows such a two-hole unit cell.  Fig.~\ref{fig:structure}(b) shows the first Brillouin zone (dashed line), which is a hexagon.  We denote the high-symmetry points~\cite{sakoda01} by $\Gamma$, $\widetilde{\rm{M}}$ and $\widetilde{\rm{K}}$. Alternatively, we can also view this structure as a triangular lattice of six-hole unit cells [white dashed hexagons in Fig.~\ref{fig:structure}(a) which we call honeycomb clusters], where the relevant parameters are the lattice constant of the triangular lattice $a_0$, the distance between the center of each cluster to the centroid of each triangular hole $R$, the length of each side of the equilateral triangular holes $s$, and the height of the dielectric material $h$. In the honeycomb lattice, the relationship $R = a_0/3$ holds.   Fig.~\ref{fig:structure}(b)  shows the first Brillouin zone as a solid hexagon and $\Gamma$, M and K indicate the high symmetry points. Note that the first Brillouin zone for the six-hole unit cell is smaller than for the two-hole unit cell due to the larger real space unit cell area. 

\begin{figure}[htbp]
\centering
\includegraphics[width=0.8\columnwidth]{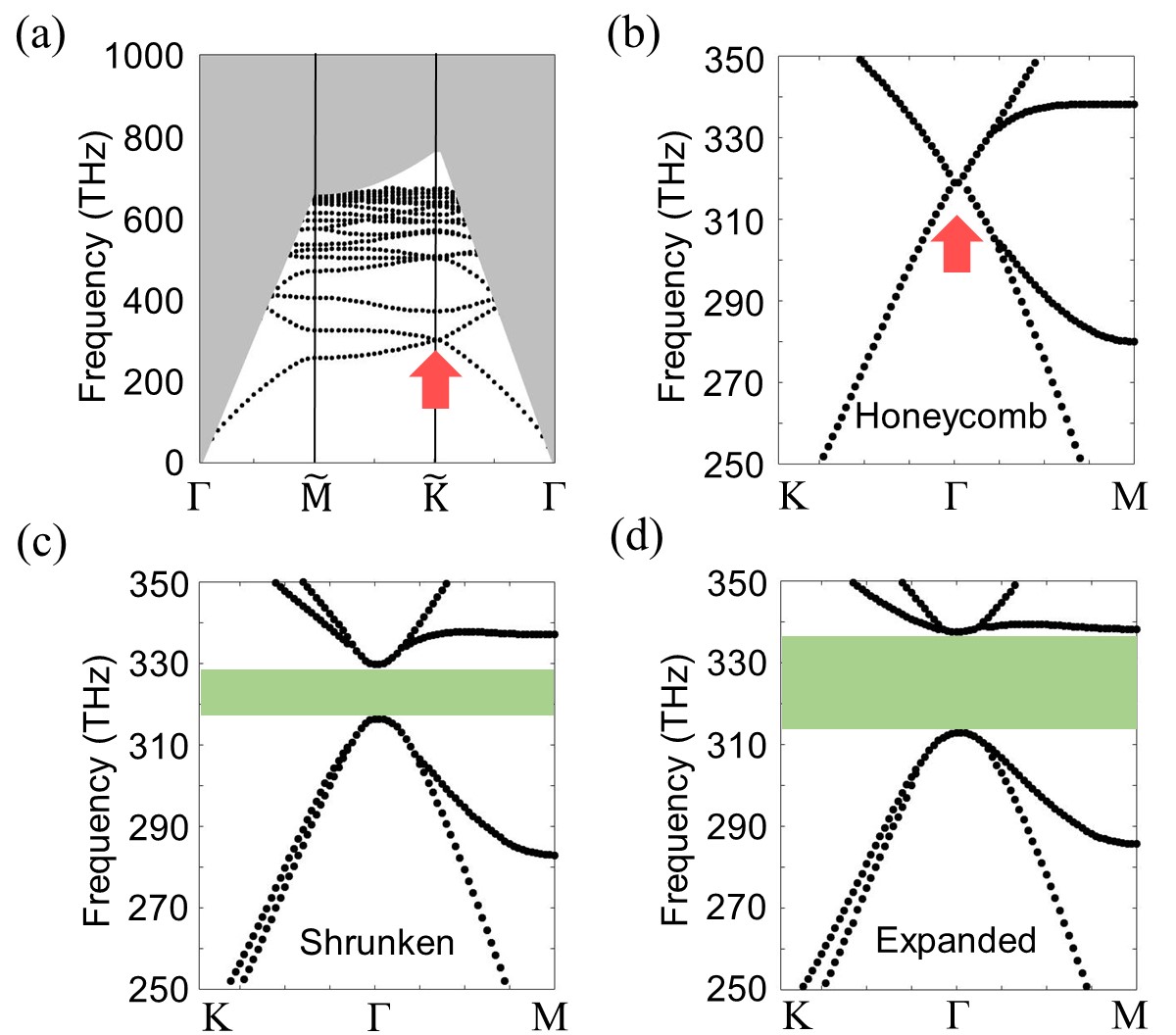}
\caption{\label{fig:band} Band structures show opening and closing of a band gap around the Dirac point as we perturb the lattice. (a) Band structure of the honeycomb lattice in the two-hole unit cell picture. The gray area represents the region above the light line, where light can leak out of the plane. A Dirac point exists at the $\widetilde{\rm{K}}$ point (red arrow) and is below the light line. (b),(c) and (d) Band structure calculated with the six-hole unit cell with honeycomb clusters ($R = a_0/3$), shrunken clusters ($R = 0.91 \times a_0/3$), and expanded clusters ($R = 1.09 \times a_0/3$) respectively. The red arrow indicates the Dirac point, and the green areas represent the band gap.}
\end{figure}

We first analyze the band structure of this photonic crystal in the two-hole unit cell picture using three-dimensional numerical finite-difference time-domain calculations (Lumerical FDTD Solutions). We perform simulations using GaAs as the dielectric substrate, with index of refraction taken from Ref.~\cite{palik97}. The parameters we use are $a_0=445$ nm, $s=140$ nm, and $h=160$ nm, which are typical dimensions for photonic crystal structures~\cite{painter99,englund05,kim13}. We focus on the transverse-electric-like modes of the system where the electric field at the symmetric plane of the system lies in-plane. Fig.~\ref{fig:band}(a) shows the band structure of the honeycomb lattice corresponding to Fig.~\ref{fig:structure}(a) along the high-symmetry points of the Brillouin zone. The gray region indicates the portion of the band structure above the light line where there are no guided modes confined in the dielectric material of finite thickness~\cite{johnson99}.  There is a Dirac point at the $\widetilde{\rm{K}}$ point, indicated by the red arrow in Fig.~\ref{fig:band}(a), located below the light line. Near this Dirac point, we can modify the topological properties of the photonic crystal by changing the ratio $R/a_0$~\cite{wu15}. However, these perturbations also change the symmetry of the lattice and so we can no longer use the rhombus-shaped two-hole unit cell to construct the band structure. Instead, we use the hexagonal six-hole unit cell to construct the band structure without destroying the rotational symmetry of the system.

We obtain the band structure for the six-hole unit cell by appropriate band folding of the bands obtained from the two-hole unit cell~\cite{suppl}. Although both Brillouin zones share the same $\Gamma$ point, the $\widetilde{\rm{K}}$ and $\widetilde{\rm{K}}'$ points for the two-hole unit cell~\cite{neto09} become folded over onto the $\Gamma$ point of the six-hole unit cell to form a doubly-degenerate Dirac point at 319 THz (which corresponds to 940 nm) as indicated by the red arrow in Fig.~\ref{fig:band}(b).

We perturb this system by varying $R$ with respect to $a_0$ to get clusters that are shrunken ($R<a_0/3$) or expanded ($R>a_0/3$) as shown in Figs.~\ref{fig:structure}(c) and (d) respectively. Figs.~\ref{fig:band}(c) and (d) show the corresponding band structures specifically for $R = 0.91 \times a_0/3$ and $R = 1.09 \times a_0/3$ respectively. Increasing or decreasing the ratio $R/a_0$ about the honeycomb lattice opens a band gap at the Dirac point. In particular, the band gaps are 13 THz and 25 THz wide for the shrunken and expanded clusters respectively. By comparing the eigenstates at the $\Gamma$ point for the expanded and shrunken structures, we see that the eigenstates are inverted between the two structures, indicating that the band topology changes as we tune the ratio $R/a_0$~\cite{suppl}.

To further confirm the numerically observed band inversion, we also analytically study the system with a tight-binding model~\cite{suppl}. The Hamiltonian of our system reduces to the Bernevig-Hughes-Zhang model for the quantum spin Hall effect~\cite{bernevig06}, where the mass term changes sign when the clusters are shrunken and expanded around $R=a_0/3$.  Consequently, the bands acquire non-zero Chern numbers that are the direct indication of non-trivial band topology. In this case, the polarization profile of the in-plane electric field acts as the pseudo-spin~\cite{suppl}.

\begin{figure}[htbp]
\centering
\includegraphics[width=0.8\columnwidth]{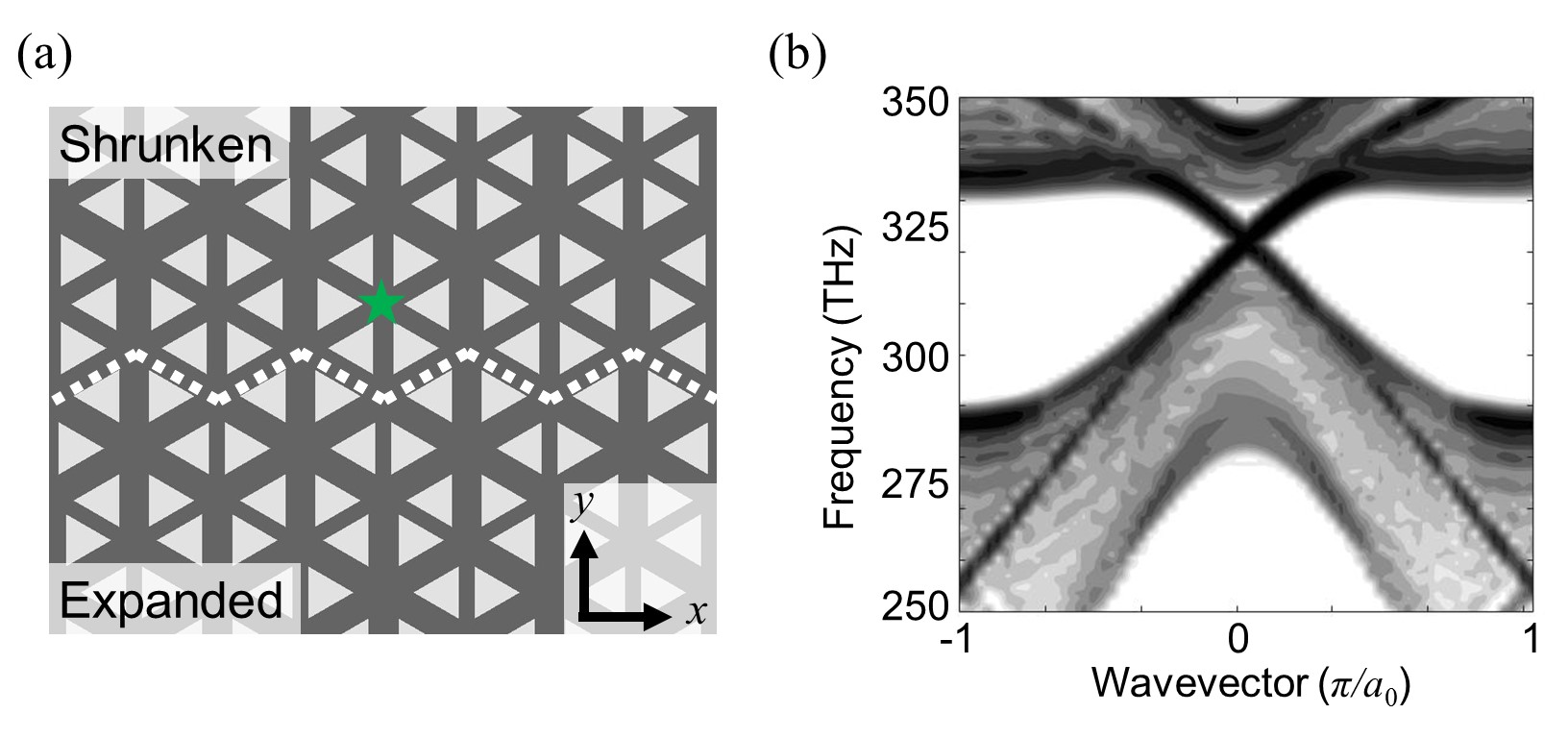}
\caption{\label{fig:edge}
Schematic and band structure which gives rise to topological edge states. (a) Schematic of two regions with different band topologies. White dotted line marks the boundary between the two regions. The star (green) indicates the location where we placed a circularly-polarized electric dipole to excite topological edge states. (b) Corresponding one-dimensional band structure shows two bands crossing the band gap in bulk. The opposite group velocities in the crossing region indicate the existence of counter-propagating directional edge states.}
\end{figure}

\section{Two-Dimensionally Confined Topological Edge States}
Non-trivial band topologies manifest themselves most dramatically in the form of guided topological edge states at the boundary between two gapped regions that have different band topologies. To confirm this, we perform three-dimensional simulations of the structure schematically shown in Fig.~\ref{fig:edge}(a) using the same values for the parameters $a_0$, $s$ and $h$ as previously. We examine topological edge states at an interface between one region composed of unit cells with shrunken clusters (13 clusters wide) and  another region of expanded clusters (12 clusters wide). These two regions share a common band gap in bulk as shown in Figs.~\ref{fig:band}(c) and (d).

Figure~\ref{fig:edge}(b) shows the one-dimensional band structure along the $x$-direction. Note that introducing an interface creates two bands crossing the original bandgap of the individual regions. The two newly formed bands have opposite group velocities, indicating counter-propagating directional edge states.

\begin{figure}[htbp]
\centering
\includegraphics[width=0.75\columnwidth]{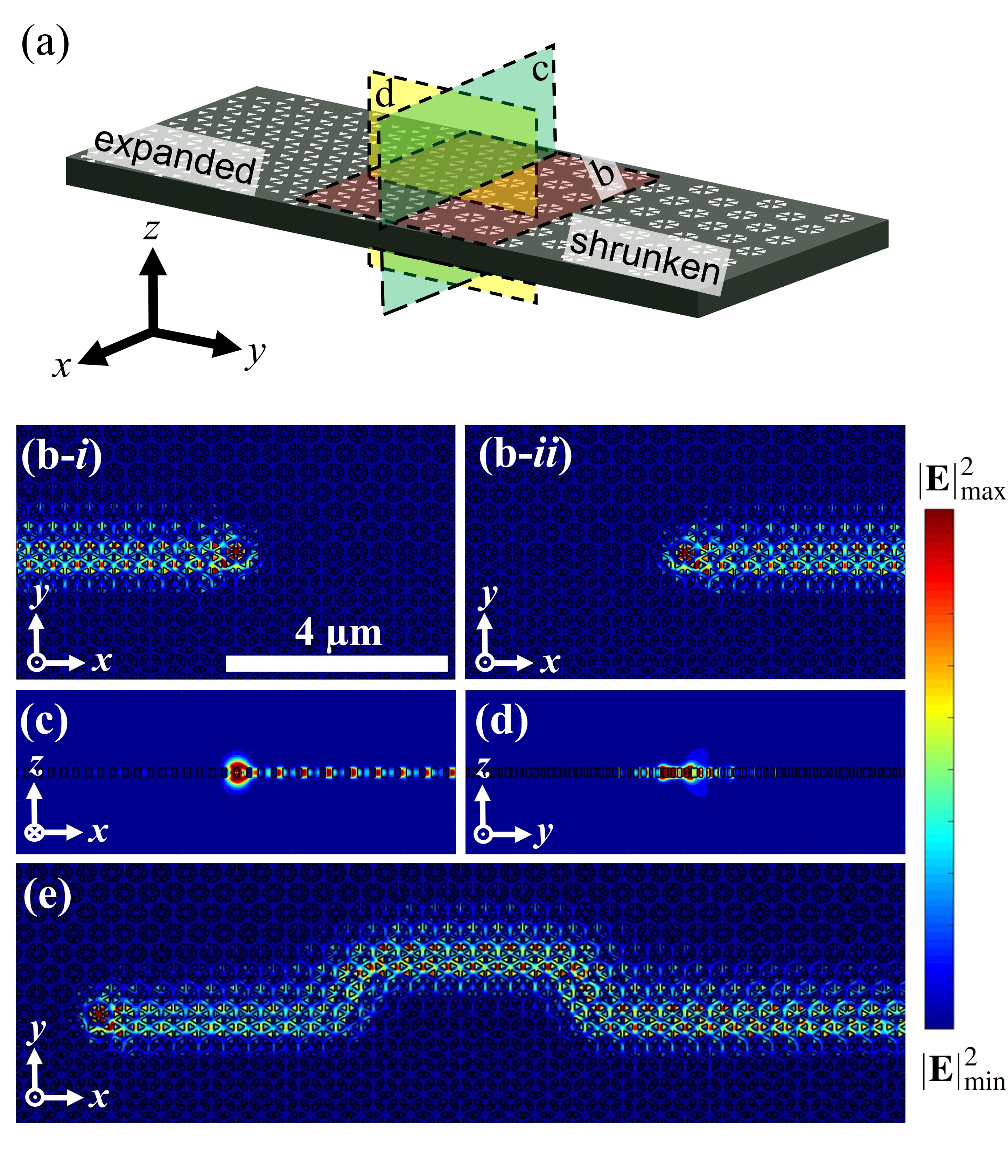}
\caption{\label{fig:3d} Three-dimensional, vertically-confined topological edge states at optical frequencies in an all-dielectric material. (a) Schematic diagram of the three-dimensional photonic crystal where the colored planes correspond to the cross-sections shown in (b), (c), and (d). (b-$i$) and (b-$ii$) Electric field intensities for a topological edge state excited with a positively and negatively circularly-polarized electric dipole show directional propagation in the $-x$ and $+x$ directions respectively. (c) and (d) Cross-section view along the $xz$ and $yz$ plane of the electric field intensity confirms that total internal reflection at the air-dielectric boundary prevents light from leaking out of the plane. (e) Electric field intensity for an edge state with four $90^{\circ}$ bends show that light can propagate around defects without backscattering.}
\end{figure}

The edge states in this system are helical, \textit{i.e.}, the pseudo-spin degree of freedom controls the direction of propagation~\cite{hasan10}.  We verify the helicity of the edge states by exciting the system with a circularly-polarized electric dipole placed at the location indicated by the green star in Fig.~\ref{fig:edge}(a). By choosing the excitation polarization to be positively (negatively) circularly polarized, we can selectively excite an edge mode propagating in the $-x$ ($+x$) direction (Fig.~\ref{fig:3d}(b-$i$)[(b-$ii$)]). The excitation frequency is 320 THz (equivalent to a wavelength of 938 nm).

Figs.~\ref{fig:3d}(c) and (d) show the electric field intensity distribution of the three-dimensional, vertically confined edge state [corresponding to Fig.~\ref{fig:3d}(b-$ii$)] in $xz$ and $yz$ cross-sections respectively. The field is confined within the dielectric slab due to total internal reflection at the air-dielectric boundary. This proves that one can realize topological edge states in three dimensions within dielectric materials at optical frequencies without significant out-of-plane loss.

One of the most distinguishing features of topological edge states is their robustness against perturbations. To test this robustness, we introduced four $90^{\circ}$ bends to the structure as shown in Fig.~\ref{fig:3d}(e). Excitation of the edge mode in this configuration shows that there is very little backscattering along the entire path. Thus our edge states exhibit topological protection against certain disorder and defects, in contrast to chiral, but topologically-trivial, waveguide modes~\cite{petersen14,sollner15}.

\section{Discussion and Conclusion}

We note that the topological protection we obtain in the presence of time-reversal symmetry differs in an important respect from that of electronic quantum spin Hall systems. The general classification of topological insulators reveals that the $\mathbb{Z}_2$ topological invariant describing the latter requires that $T^2 = -1$, where $T$ is the time-reversal symmetry operator. The minus sign is a particular feature of fermionic systems. In contrast, Maxwell's equations (and other bosonic systems) obey $T^2 = 1$. This symmetry taken alone does not afford any topological protection in two dimensions. 

However, we can construct a `pseudo' time-reversal symmetry operator based on the ($C_{6v}$) crystal symmetry of the lattice which obeys $T^2 = -1$~\cite{wu15}. While this assures that the bulk may be classified according to a $\mathbb{Z}_2$ topological invariant, gapless edge modes are not guaranteed since this symmetry is broken at the boundaries. This symmetry breaking can mix the counter-propagating edge states and open a mini-gap in the edge mode~\cite{longphotonic}; in a quantum spin Hall system, this would be akin to a magnetic impurity at the edge of the system. Apparently, in our realization this symmetry breaking is weak since we do not observe a gap in the edge states [Fig~\ref{fig:edge}(b)]. We can decouple the pseudo-spin degrees of freedom up to linear order in $\mathbf{k}$ near the $\Gamma$ point. By considering these degrees of freedom as being completely decoupled, we can characterize the topology of the system by a stronger $\mathbb{Z}$ spin Chern number given by the difference of the Chern numbers for each pseudo-spin~\cite{albert15}.

To conclude, we have proposed a new all-dielectric photonic crystal design and presented simulation results showing that three-dimensionally guided topological edge states at optical frequencies can be realized. Our design parameters are amenable to implementation with well-established nanofabrication techniques. Our simulations focus on GaAs as  the dielectric substrate but the photonic crystal design principles that give rise to topological edge states are applicable to many other dielectric materials such as indium phosphide, silicon, and diamond. With the future prospect of integration with various quantum emitters ranging from quantum dots~\cite{press08,berezovsky08,sun16}, defects in two-dimensional materials~\cite{xia14} and diamond~\cite{aharonovich11,aharonovich11b}, this system promises to open a new path to research in topological phenomena with optical systems.

\section{Acknowledgments}

We thank Sunil Mittal, Mikael Rechtsman, Alberto Amo, and Jay Sau for fruitful discussions. This research was supported by Air Force Office of Scientific Research, Sloan Foundation, Office of Naval Research and the Physics Frontier Center at the Joint Quantum Institute. KITP is supported by NSF PHY11-25915.

\section{Appendix}


\subsection{Band Structure of a Honeycomb Lattice with Circular Holes}

\begin{figure}[h]
\begin{center}
\includegraphics[width=0.7\columnwidth]{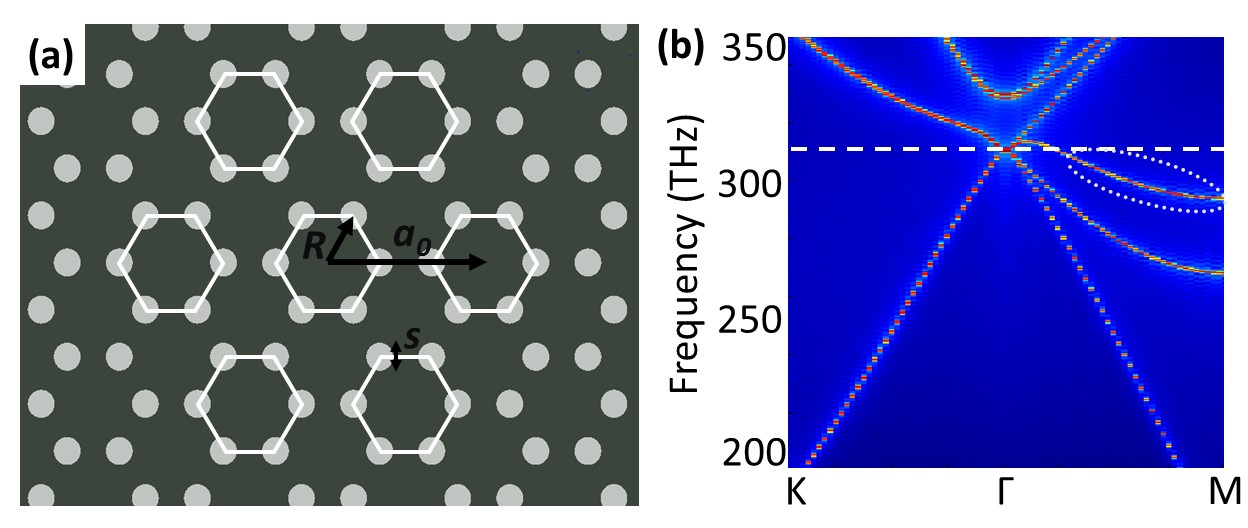}
\caption{Band structure of a honeycomb lattice with circular holes. (a) Schematic of a honeycomb lattice made of circular holes, where the parameters $a_0$ is the lattice constant of the hexagonal clusters (white hexagons) which constitute a triangular lattice, $R$ is the distance from the center of the cluster to the center of a circular hole within the cluster ($R = a_0/3$ in this case), and $s$ is the diameter of the circular hole. (b) Band structure for transverse-electric modes of the structure shown in (a), showing the appearance of a Dirac cone at 312 THz (indicated by the horizontal white dashed line). White dotted ellipse shows one of the bands crossing the frequency at the Dirac point, which prevents the appearance of a band gap across the Brillouin zone after perturbation. Calculations were done with $a_0 = 350$ nm and $s = 140$ nm.}
\label{fig:circle_band}
\end{center}
\end{figure}

One idea is to implement our honeycomb-lattice-like photonic crystal structure with circular holes instead of triangular holes, as shown in Fig.~\ref{fig:circle_band}(a). Although the band structure of the transverse-electric modes, as shown in Fig.~\ref{fig:circle_band}(b), does give rise to a Dirac point, it turns out that it does not give rise to a band gap in the region of interest. The horizontal white dashed line shows the frequency at the Dirac point and the white dotted region encloses a range of wavevectors for which one of the bands crosses the frequency at the Dirac point, thus preventing the appearance of a band gap across the Brillouin zone even after perturbation, which is critical for realizing topological edge states. This can be avoided with the use of equilateral triangular holes, where a band gap is possible after perturbations to the system.

\subsection{Honeycomb Lattice with a Six-Site Basis and Band Folding}

\begin{figure}
\begin{center}
\includegraphics[width=0.9\columnwidth]{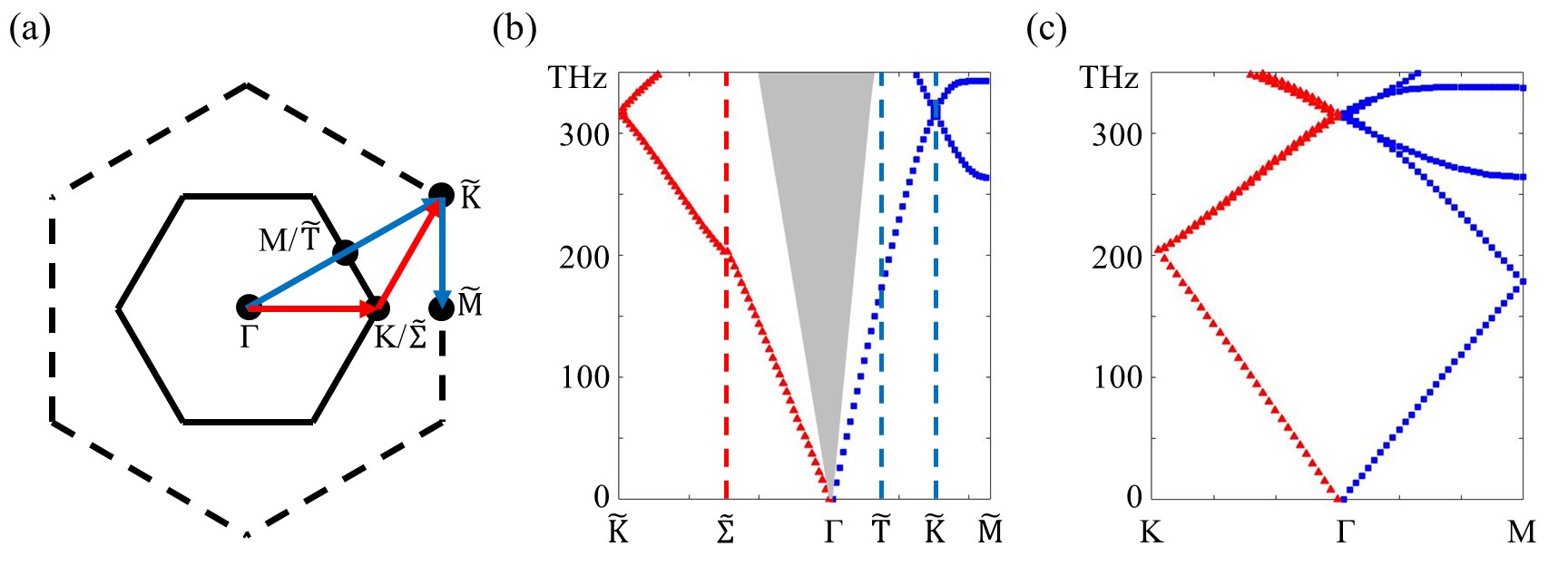}
\caption{Correspondence between the two-site and six-site bases. (a) Boundaries of the first Brillouin zone for the two-site (dashed hexagon) and six-site (solid hexagon) bases for the honeycomb lattice. (b) The band structure of the honeycomb lattice considered with a two-site basis. The labels on the horizontal axis corresponds to the high-symmetry points in $k$-space as designated in (a)~\cite{sakoda01}. The gray area is the area above the light cone where guided modes are not possible. The red triangles (blue squares) correspond to the red (blue) paths indicated in (a). (c) The band structure of the honeycomb lattice with a six-site basis as obtained by folding the band structure in (b) along the dotted vertical lines.
\label{fig:band_folding}}
\end{center}
\end{figure}

The photonic crystal we study is a modification of the usual honeycomb lattice. For the special case that the lattice parameters obey $R = a_0/3$, the standard honeycomb lattice is recovered (see main text for definitions). Typically, the honeycomb lattice is taken to be a triangular lattice with a two-site basis. For the general case $R \neq a_0/3$, it is convenient to consider the system as a triangular lattice with a six-site basis with primitive lattice vectors \begin{eqnarray}
\mathbf{a}_1 &=& \left( \sqrt{3}, 0 \right) a, \\
\mathbf{a}_2 &=&  \left( \sqrt{3}/2, 3/2 \right) a,
\end{eqnarray}
where $a = \sqrt{3}R$. Figure~\ref{fig:band_folding}(a) shows the first Brillouin zones (FBZs) for both the two-site (dashed hexagon) and six-site (solid hexagon) bases.

The equivalence of these two descriptions can be verified by counting the total number of states in each case. The two-site basis is described by two bands over the FBZ of area $A$, giving a total number of states corresponding to an effective $k$-space area of $2A$. In the six-site case, each linear dimension of the FBZ is reduced by a factor $\sqrt{3}$ and thus the area is $A/3$. Since there are six bands, this again gives a total effective area of $6 \times A/3 = 2A$.

In the case of graphene, it is well-known that the Dirac cones are located at the edges of the FBZ [labeled by $\widetilde{K}$ in Fig.~\ref{fig:band_folding}(a)]. In the six-site basis, these degrees of freedom now reside at the zone center ($\Gamma$). The $\widetilde{K}$ and $\Gamma$ points are connected by a reciprocal lattice vector. The bands in Fig.~\ref{fig:band_folding}(c) can be obtained by folding along the vertical dashed lines in Fig.~\ref{fig:band_folding}(b) so that the $\widetilde{K}$ is matched to $\Gamma$. At this point, the bands formerly at the two inequivalent Dirac points $\widetilde{K}$ and $\widetilde{K}'$ come together to form a doubly degenerate Dirac point. We will designate these degrees of freedom by a pseudospin ($\pm$)~\cite{wu15}.

\subsection{Tight-Binding Description of the Dispersion}
 
The dispersion of our system near $\Gamma$ ($\textbf{k} = 0$) can be obtained by a tight-binding model. In this approach, we take a set of basis states for which the magnetic field profile is concentrated in a particular hole. The time-evolution of the system is characterized by `hopping' to adjacent holes in the lattice. For the case of a standard honeycomb lattice (i.e. $a_0 = 3R$), symmetry dictates that the amplitude to hop to an adjacent lattice site is the same whether that site is a member of the same unit cell or a neighboring one. For this geometry, we find that dispersion remains gapless at the Dirac point. For $a_0 \neq 3R$, a gap develops at the Dirac points which is proportional to the detuning. After presenting this analysis, we will discuss the applicability of this method to the photonic system.

\begin{figure}
\begin{center}
\includegraphics[width=0.7\columnwidth]{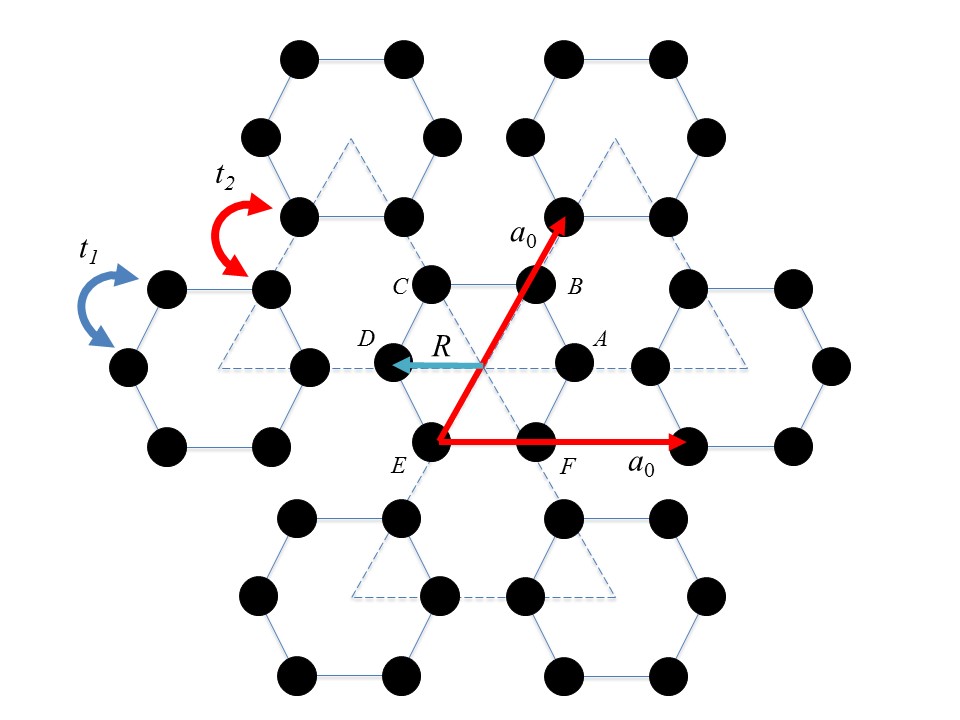}
\caption{Schematic of our lattice parameters and the labeling of the lattice sites for our tight-binding model. A cluster consists of six sites. Then the system is a triangular lattice of clusters with lattice constant $a_0$.  The distance from the centroid of each hole to the center of its cluster is $R$. The tunneling amplitudes $t_1$ and $t_2$ correspond to intra- and inter-cluster tunneling between the nearest neighbor holes. The labels {\it{A, B, C, D, E, F}} denote each lattice site making up the basis.}
\label{fig:honeycomb_schematic}
\end{center}
\end{figure}

We describe our system as a triangular lattice with a six-site basis labeled $A, B, C..., F$ starting with the right-most site and progressing in a counter-clockwise manner (Fig.~\ref{fig:honeycomb_schematic}). The states of our system $|A \rangle$, $|B \rangle$, $|C \rangle$...,$|F \rangle$ are the Wannier functions for the system. For example, the state $|C \rangle$ describes an electromagnetic field configuration for which the out-of-plane magnetic field is centered on the $C$ hole in each six-membered ring. In the bandwidth of interest, the magnetic field configurations can be written as linear combinations
\begin{equation}
|\Psi \rangle = \left(
\begin{array}{c}
  \psi_A \\
  \psi_B \\
  \psi_C \\
  \psi_D \\
  \psi_E \\
  \psi_F
\end{array}\right) e^{i \mathbf{k} \cdot \mathbf{r}},
\label{eq:Hz}
\end{equation}
where $\mathbf{r} = (x,y)$ and Eq.~(\ref{eq:Hz}) is written in the basis
\begin{equation}
|A\rangle =
\left(
\begin{array}{c}
  1 \\
  0 \\
  0 \\
  0 \\
  0 \\
  0
\end{array}\right), \
|B \rangle =
\left(
\begin{array}{c}
  0 \\
  1 \\
  0 \\
  0 \\
  0 \\
  0
\end{array}\right),... \
|F \rangle =
\left(
\begin{array}{c}
  0 \\
  0 \\
  0 \\
  0 \\
  0 \\
  1
\end{array}\right).
\end{equation}

The action of the Hamiltonian operator is to evolve the state in time. Roughly, the matrix elements of $\mathcal{H}$ indicate the field configurations which can evolve into each other on a time scale $\sim R / c$. On this time scale, only states which are localized to adjacent sites can evolve into each other appreciably and thus we only consider nearest-neighbor `hopping'. The Hamiltonian $\mathcal{H} = \mathcal{H}_1 + \mathcal{H}_2$ receives contributions from intra- and inter-cluster couplings, respectively. Intra-cluster hopping is characterized by a parameter $t_1$ and takes the form
\begin{equation}
\mathcal{H}_1 = - t_1 \left( \begin{array}{cccccc}
             0 & 1 & 0 & 0 & 0 & 1 \\
             1 & 0 & 1 & 0 & 0 & 0 \\
             0 & 1 & 0 & 1 & 0 & 0 \\
             0 & 0 & 1 & 0 & 1 & 0 \\
             0 & 0 & 0 & 1 & 0 & 1 \\
             1 & 0 & 0 & 0 & 1 & 0
           \end{array}
 \right).
 \label{eq:Ham1}
\end{equation}
Inter-cluster coupling is described by
\begin{equation}
\mathcal{H}_2 = - t_2 \left( \begin{array}{cccccc}
             0 & 0 & 0 & e^{i \mathbf{k} \cdot \mathbf{a}_1} & 0 & 0 \\
             0 & 0 & 0 & 0 & e^{i \mathbf{k} \cdot \mathbf{a}_2 } & 0 \\
             0 & 0 & 0 & 0 & 0 & e^{ i \mathbf{k} \cdot \left( \mathbf{a}_2 - \mathbf{a}_1 \right)} \\
             e^{- i \mathbf{k} \cdot  \mathbf{a}_1} & 0 & 0 & 0 & 0 & 0 \\
             0 &  e^{-i \mathbf{k} \cdot \mathbf{a}_2} & 0 & 0 & 0 & 0 \\
             0 & 0 & t_2 e^{- i \mathbf{k} \cdot \left( \mathbf{a}_2 - \mathbf{a}_1 \right)} & 0 & 0 & 0 
\end{array}
 \right).
\label{eq:Ham2}
\end{equation}

We begin by diagonalizing at $\mathbf{k} = (0,0)$. We consider the ans\"{a}tz
\begin{equation}
| \Psi(z) \rangle = \frac{1}{\sqrt{6}}\left(
\begin{array}{c}
  1 \\
  z \\
  z^2 \\
  z^3 \\
  z^4 \\
  z^5
\end{array}\right),
\end{equation}
where $z$ is a complex number (note it is not position) with $|z| = 1$. As we will see, the phase of $z$ is analogous to an angular momentum. Acting $\mathcal{H}$ on $|\Psi(z) \rangle$ yields
\begin{equation}
\mathcal{H} | \Psi(z) \rangle = \left(
\begin{array}{c}
  -t_1 \left( z^5 + z \right) - t_2 z^3 \\
  -t_1 \left( 1   + z^2 \right) - t_2 z^3 \\
  -t_1 \left( z   + z^3 \right) - t_2 z^3 \\
  -t_1 \left( z^2 + z^4 \right) - t_2 z^{-3} \\
  -t_1 \left( z^3 + z^5 \right) - t_2 z^{-3} \\
  -t_1 \left( z^4 + 1 \right) - t_2 z^{-3}
\end{array}\right).
\end{equation}
We find that $| \Psi(z) \rangle$ is an eigenstate provided that $z$ is a sixth root of unity. For $z = e^{i \pi n /3}$, the spectrum is given by
\begin{equation}
\varepsilon_n = -2 t_1 \cos \frac{\pi n}{3} - (-1)^n t_2.
\end{equation}
Thus, we see that the $|\Psi(z) \rangle$ states associated with $z = 1$ and $z = -1$ have energies $-2t_1 - t_2$ and $2 t_1 + t_2$, respectively. Time-reversal symmetry is enacted by complex conjugation. Thus the states corresponding to $z = e^{i \pi / 3}$ and $e^{- i \pi / 3}$ have energies $\varepsilon_{\pm1} = -t_1 + t_2$ while $z = e^{i 2 \pi / 3}$ and $e^{- i 2 \pi / 3}$ have energies $\varepsilon_{\pm2} = t_1 - t_2$.

Although the full rotational symmetry is broken by the crystal axis, the states corresponding to $z = e^{\pm i \pi / 3}$ possess strong $p$-like character, while those with $z = e^{\pm i 2 \pi / 3}$ have $d$-like character. This can be most easily seen by noting that the various states $|\Psi(z) \rangle$ are `sampled' from continuous angular wave functions as follows 
\begin{eqnarray}
| \Psi(e^{\pm i \pi / 3}) \rangle = e^{ \pm i \theta} \rightarrow | p_\pm \rangle, \\
| \Psi(e^{\pm i 2 \pi / 3}) \rangle =  e^{\pm i 2 \theta} \rightarrow |d_\pm \rangle.
\end{eqnarray}
The $\pm$ labels the pseudo-spin degree of freedom. The geometry of the wavefunctions is clarified through the definitions
\begin{eqnarray}
| p_x \rangle &=& \frac{1}{\sqrt{2}} \left( | p_+ \rangle + | p_- \rangle  \right), \\
| p_y \rangle &=& \frac{1}{i \sqrt{2}} \left( | p_+ \rangle - | p_- \rangle  \right)
\end{eqnarray}
where $|p_x \rangle$ is odd about the $x$-axis, etc. Similarly, we have
\begin{eqnarray}
| d_{x^2 - y^2} \rangle &=& \frac{1}{\sqrt{2}} \left( |d_+\rangle + |d_-\rangle \right), \\
| d_{xy} \rangle &=& \frac{1}{i \sqrt{2}} \left( |d_+ \rangle - |d_-\rangle \right),
\end{eqnarray}
where $| d_{x^2 - y^2} \rangle$ is a wave function whose maxima coincide with the $x$- and $y$-axes as $\theta = 0 \rightarrow 2\pi$, etc.

We now derive the spectra associated with these 4 states near $\Gamma$ by expanding Eq.~(\ref{eq:Ham2}) to linear order in $k_x$ and $k_y$. In this limit, the effective $4 \times 4$ Hamiltonian is block diagonal, and only states of the same pseudo-spin are coupled. The effective Hamiltonian for the ($+$)-pseudo-spin is given by
\begin{equation}
\mathcal{H}_+ = \frac{\sqrt{3}}{2} t_2 a \left( - k_x \sigma_x + k_y \sigma_y \right) + \left[ t_2 - t_1 + \mathcal{O}(k_x^2+k_y^2)\right] \sigma_z,
\label{eq:Hplus}
\end{equation}
in the $(|p_+\rangle, |d_+\rangle)^T$ basis. Similarly, in the $(|p_-\rangle,|d_-\rangle)^T$ basis we find
\begin{equation}
\mathcal{H}_- = \frac{\sqrt{3}}{2} t_2 a \left( k_x \sigma_x + k_y \sigma_y \right) + \left[ t_2 - t_1 + \mathcal{O} (k_x^2+k_y^2) \right] \sigma_z.
\label{eq:Hminus}
\end{equation}
In both cases, we have performed a unitary transformation $U = e^{ i \frac{\pi}{2} \sigma_z}$. We note that in the limit that the various honeycombs are completely decoupled, $t_2 \approx 0$ and Eqs.~(\ref{eq:Hplus}) and (\ref{eq:Hminus}) reflect the fact that the $p$-states have a lower energy than the $d$-states. For $t_1 = t_2$, $\mathcal{H}_+$ and $\mathcal{H}_-$ are characterized by a Dirac cone spectrum. For $t_1 \neq t_2$, the spectrum acquires a gap of size $|t_1 - t_2|$.

Typically, the application of tight-binding is limited to electronic systems in which electrons hop between weakly coupled atomic orbitals. However, the method is actually much more broadly applicable. It turns out that \emph{any} band can always be written in terms of so-called \emph{Wannier functions}~\cite{ashcroft}. Only if the various `atomic' states are weakly coupled will the Wannier functions bear a strong resemblance to the atomic wave functions, but generally such Wannier functions may always be obtained. Moreover, the band structure near $\Gamma$ is tightly constrained by the symmetries of the system. In particular, the tight-binding Hamiltonian $\mathcal{H}$ automatically accounts for the fact that the lattice and the triangular holes exhibit a $C_{6v}$ symmetry. For $t_1 = t_2$, the Dirac cones are protected by additional $C_{3v}$ symmetries.

\subsection{Topology and Edge States} 

In the previous section, we showed that a honeycomb structure can be described by a gapless Dirac Hamiltonian. When we introduce the lattice deformations, \textit{i.e.}, shrinking/expanding, a gap opens which can be described a mass term ($m\sigma_z$). Here, we review the concept why the band inversion, \textit{i.e.}, changing the sign of mass, results in having a topological edge at the boundary. 

When the system is gapped, its topology can be characterized by a Chern number for the pseudospins ($\pm$). A spin Chern number takes the form
\begin{equation}
\mathcal{C} = \mathcal{C}_+ - \mathcal{C}_-,
\end{equation}
where $\mathcal{C}_\pm = \pm \frac{1}{2}$sgn$(m_\pm)$, where $m_\pm$ are the masses for the two pseudo-spins~\cite{haldane}. Thus, we have 
\begin{equation}
\mathcal{C} = \textrm{sgn}(t_2 - t_1).
\end{equation}
Topologically-protected edge modes will exist between gapped regions with different $\mathcal{C}'s$, \textit{i.e.}, any place that the quantity $t_2 - t_1$ changes sign.

In order to understand the edge state structure, we begin by considering $\mathcal{H}_+$ with a spatially varying mass. For concreteness, we consider the situation outlined in Fig. 4b in the main text  As we will see, edge states are localized to domain walls for which $m(x) = t_2 - t_1 \approx 0$. The edge states satisfy the Heisenberg equation of motion which, for $\mathcal{H}_+$ [Eq.~(\ref{eq:Hplus})], is the Dirac equation.  The Dirac equation corresponding to $\mathcal{H}_+$ is
\begin{equation}
 \left[ - i \hbar v \left( - \sigma_x \partial_x + \sigma_y \partial_y \right) + m \sigma_z \right] \Psi = E \Psi,
\end{equation}
where $v = \sqrt{3} t_1 a_0 / 2$ and $E$ is the energy of the eigenstate $\Psi$.

Consider the geometry shown in Fig. 4(b) of the main text, which shows an area of shrunken hexagons above expanded hegaons. The system is described by a mass which depends only on $y$, \textit{i.e.}, $m(x,y) = m(y)$ and $m(0) = 0$ with
\begin{equation}
\frac{dm}{dy} < 0.
\end{equation}
In this case, the topologically protected solution 
\begin{equation}
\Psi(y) = \chi \exp \left( \frac{1}{\hbar v} \int_0^y m(y') dy' \right), \label{eq:psiborder}
\end{equation}
is an $x$-independent solution of the Dirac equation with zero energy where $\chi$ is a two-dimensional spinor. This is the celebrated Jackiw-Rebbi solution of the Dirac equation with a spatially varying mass~\cite{jackiw}. The sign in the exponent of $\Psi$ [Eq.~(\ref{eq:psiborder})] ensures that the solution is normalizable. The edge state decays exponentially for both $y > 0$ and $y<0$. The spinor $\chi$ obeys
\begin{equation}
\sigma_x \chi = \chi.
\end{equation}
Thus,
\begin{equation}
\chi = \frac{1}{\sqrt{2}} \left(
                            \begin{array}{c}
                              1 \\
                              1 \\
                            \end{array}
                          \right),
\end{equation}
in the $(|p_+\rangle, |d_+\rangle)^T$ basis. The full edge mode is described by
\begin{equation}
\Psi(x,y) = \frac{1}{\sqrt{2}} \left(
                            \begin{array}{c}
                              1 \\
                             1 \\
                            \end{array}
                          \right) \exp \left( \frac{1}{\hbar v} \int_0^y m(y') dy' \right) e^{i k_x x}.
\end{equation}
Again, plugging into the Dirac equation gives an energy dispersion
\begin{equation}
E(k_x) = - \hbar v k_x.
\end{equation}
Since the group velocity is given by $v = \frac{1}{\hbar} \frac{\partial E}{\partial k_x}$,
this represents an edge state travelling in the $-x$-direction. Indeed, we see that in Fig. 4(b-\textit{i}) of the main text, the excitation of the $+$-pseudospin leads to a left-moving edge state. Similarly, an edge state derived from the $\mathcal{H}_-$ channel (opposite pseudo-spin) would travel in the $+x$-direction.

\subsection{Inversion of the Eigenstates}

\begin{figure}
\begin{center}
\includegraphics[width=0.6\columnwidth]{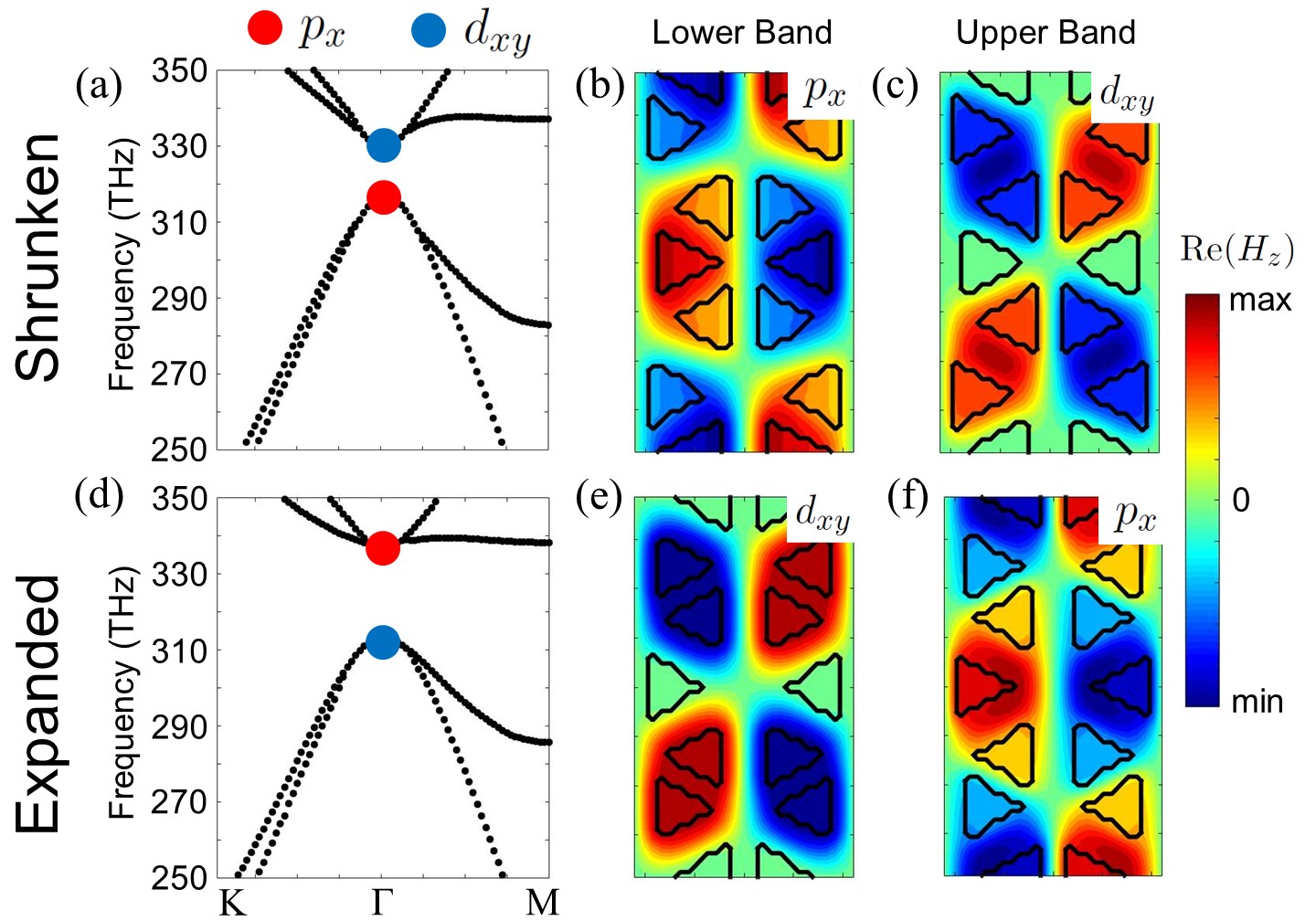}
\caption{Band inversion. (a) and (d) Band structures for the shrunken and expanded cluster systems, which are the same as Fig. 2(c) and (d) respectively in the main text, with a subset of the eigenstates indicated at the $\Gamma$ point. (b) and (c) [(e) and (f)] Out-of-plane magnetic field eigenstates at the symmetry plane $z=0$ of the lower and upper band for the shrunken (expanded) cluster system respectively. We see that $e.g.$, the eigenstate $p_x$ for the lower band in the shrunken cluster system appears on the upper band for the expanded cluster system, which indicates a change in the band topology.}
\label{fig:bandinv}
\end{center}
\end{figure}

We examine the out-of-plane magnetic field eigenstates of the system at the symmetry plane ($z=0$) corresponding to the $\Gamma$ point for the shrunken and expanded clusters. The band structures for the shrunken and expanded cluster systems are shown in Fig.~\ref{fig:bandinv}(a) and (d) and are the same as Fig. 2(c) and (e) in the main text. The eigenstates corresponding to these band structures show that the eigenstates are inverted; by that we mean that $e.g.$, the eigenstate $p_x$ ($d_{xy}$) shown in Fig.~\ref{fig:bandinv}(b) [\ref{fig:bandinv}(c)] which appeared on the lower (upper) band for the shrunken cluster appears on the upper (lower) band for the expanded cluster as shown in Fig.~\ref{fig:bandinv}(f) [(e)]. This band inversion indicates that there is a change in the band topology, as discussed in the previous section on the tight-binding model.

\subsection{Polarization Pseudo-Spin of the Eigenstates}

\begin{figure}
\begin{center}
\includegraphics[width=1\columnwidth]{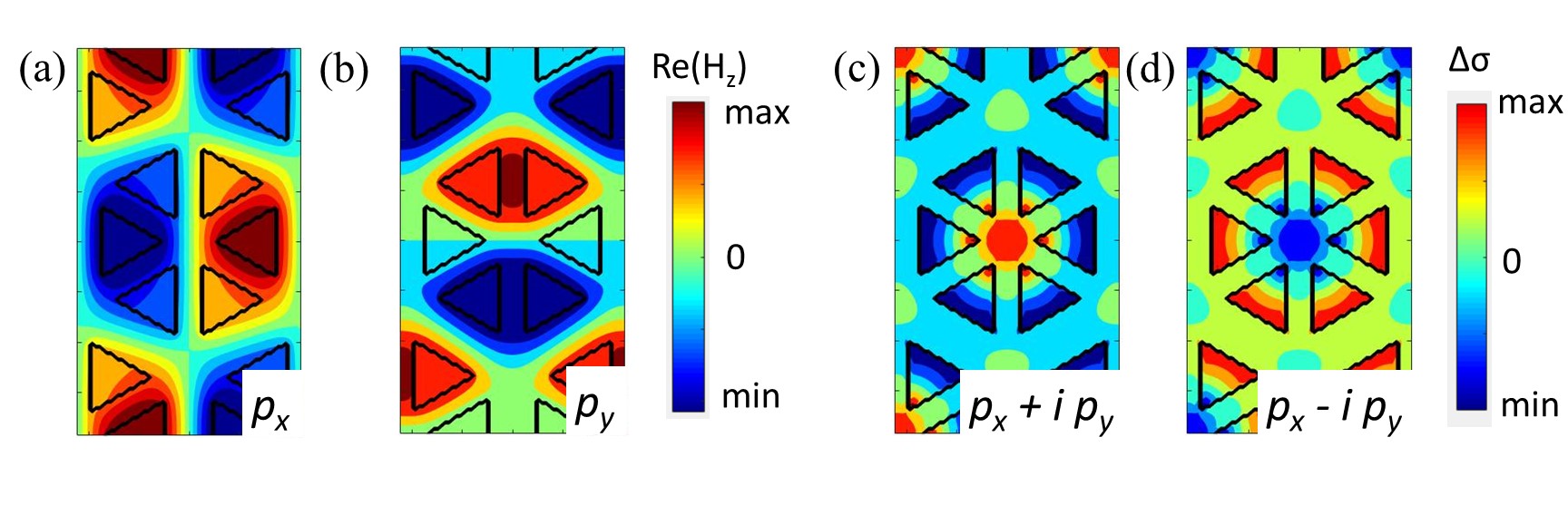}
\caption{Out-of-plane magnetic field and in-plane electric field polarization in a cluster of six holes outlined by black triangles. (a) and (b) depict the real part of the out-of-plane magnetic field ($H_z$) for the eigenstates $p_x$ and $p_y$ respectively. The colors indicate the strength of $H_{z}$. (c) and (d) show $\Delta \sigma \sim |E_+|^2 - |E_-|^2$, indicating the pseudo-spin nature of the bands excited at the $\Gamma$ point with $p_x + i p_y$ and $p_x - i p_y$ modes, respectively.}
\label{fig:polarization}
\end{center}
\end{figure}

From Maxwell's equations, at the symmetry plane $z=0$, the out-of-plane magnetic field eigenstates $p_x \hat{\textbf{z}}$ and $p_y \hat{\textbf{z}}$ lead to an in-plane electric field given by
\begin{equation}
\textbf{E}_1 = \frac{i}{\omega \epsilon_0 \epsilon(\textbf{r})} \nabla \times (p_x \hat{\textbf{z}}) \quad \qquad \textbf{E}_2 = \frac{i}{\omega \epsilon_0 \epsilon(\textbf{r})} \nabla \times (p_y \hat{\textbf{z}}),
\label{eq:e-field}
\end{equation}
where $\textbf{E}_i = E_{i x} \hat{\textbf{x}} + E_{i y} \hat{\textbf{y}}$ ($i = 1,2$), $\epsilon_0 \approx 8.854 \times 10^{-12}$ Farad/m is the vacuum permittivity, and $\epsilon(\textbf{r})$ is the position-dependent relative permittivity. The out-of-plane magnetic fields of the $p_x$ and $p_y$ eigenstates are shown in Figs.~\ref{fig:polarization}(a) and (b), respectively. We see that the $p_x$ and $p_y$ modes are related by a $\pi/2$ rotation, so we have at the center of the cluster ($\textbf{r}$ = 0) the relation

\begin{equation}
\left(\begin{array}{c}
                             E_{2x} \\
 E_{2y}  \\
                            \end{array}
                          \right) =  \left(\begin{array}{c}
                             -E_{1y} \\
                              E_{1x}  \\
                            \end{array}
                          \right)
\end{equation}

From this we find the relation
\begin{equation}
 \frac{i}{\omega \epsilon_0 \epsilon(0)} \nabla \times [(p_x \pm i p_y) \hat{\textbf{z}}] = (E_{1x} \mp i E_{1y}) (\hat{\textbf{x}} \pm i \hat{\textbf{y}}).
\end{equation}
This implies that at the center of the clusters the in-plane electric field polarization is either $\sigma_+$- or $\sigma_-$-circularly polarized depending on the out-of-plane magnetic field eigenstates $\textbf{p}_{\pm} = (p_x \pm i p_y) \hat{\textbf{z}}/\sqrt{2}$ where $\sigma_{\pm} = (\hat{\textbf{x}} \pm i \hat{\textbf{y}})/\sqrt{2}$. We can see this directly in Figs.~\ref{fig:polarization} (c) and (d), which show $\Delta \sigma \sim |E_+|^2 - |E_-|^2$ where $E_{\pm} = (E_x \mp i E_y)/\sqrt{2}$, characterizing the degree of circular polarization. In both cases we see that at the center the in-plane electric field is highly circularly polarized, except with opposite handedness. Thus the out-of-plane magnetic field eigenstates $p_{\pm}$ have an associated in-plane electric field circular polarization of $\sigma_{\pm}$ which act as pseudo-spins for this topological photonic crystal.

\section{References}


\begin{thebibliography}{99}


\bibitem{hasan10} M. Z. Hasan and C. L. Kane, Rev. Mod. Phys. \textbf{82}, 3045 (2010).
\bibitem{qi11} X.-L. Qi and S.-C. Zhang, Rev. Mod. Phys. \textbf{83}, 1057 (2011).

\bibitem{galitski13} V. Galitski and I. B. Spielman, Nature \textbf{494}, 49 (2013).
\bibitem{jotzu14} G. Jotzu \emph{et al}., Nature \textbf{515}, 237 (2014).
\bibitem{yang15} Z. Yang \emph{et al}., Phys. Rev. Lett. \textbf{114}, 114301 (2015).
\bibitem{peano15} V. Peano, C. Brendel, M. Schmidt, and F. Marquardt, Phys. Rev. X \textbf{5}, 031011 (2015).
\bibitem{wang15} P. Wang, L. Lu, and K. Bertoldi, Phys. Rev. Lett. \textbf{115}, 104302 (2015)
\bibitem{huber15}R. S\"{u}sstrunk and S. D. Huber, Science \textbf{349}, 47 (2015).

\bibitem{lu14} L. Lu, J. D. Joannopoulos, and M. Solja\v{c}i\'{c}, Nature Photon. \textbf{8}, 821 (2014).
\bibitem{hafezi14} M. Hafezi and J. M. Taylor, Physics Today  \textbf{67}, 68 (2014).

\bibitem{wang09} Z. Wang, Y. Chong, J. D. Joannopoulos, and M. Solja\v{c}i\'{c}, Nature \textbf{461}, 772 (2009).
 \bibitem{khanikev16} X. Cheng, C. Jouvaud, X. Ni, S. H. Mousavi, A. Z. Genack, and A. B. Khanikaev, Nat. Mater. \textbf{15}, 542 (2016).

\bibitem{hafezi13} M. Hafezi, S. Mittal, J. Fan, A. Migdall, and J. M. Taylor, Nature Photon. \textbf{7}, 1001 (2013).
\bibitem{mittal14} S. Mittal \emph{et al}., Phys. Rev. Lett. \textbf{113}, 087403 (2014).


\bibitem{rechtsman13} M. C. Rechtsman \emph{et al.}, Nature \textbf{496}, 196 (2013).

\bibitem{mittal16} S. Mittal, S. Ganeshan, J. Fan, A. Vaezi and M. Hafezi, Nature Photon. \textbf{10}, 180 (2016).


\bibitem{gritsev14} M. Ringel, M. Pletyukhov and V. Gritsev, New J. Phys. \textbf{16}, 113030 (2014)
\bibitem{pichler15} H. Pichler, T. Ramos, A. J. Daley, and P. Zoller, Phys. Rev. A \textbf{91}, 042116 (2015).

\bibitem{cho08} J. Cho, D. G. Angelakis, and S. Bose, Phys. Rev. Lett. \textbf{101}, 246809 (2008).
\bibitem{carusotto12} R. O. Umucallar and I. Carusotto, Phys. Rev. Lett. \textbf{108}, 206809 (2012).
\bibitem{hafezi13b} M. Hafezi, M. D. Lukin, and J. M. Taylor,  New J. Phys. \textbf{15}, 063001 (2013).

\bibitem{joannopoulos08} J. D. Joannopoulos, S. G. Johnson, J. N. Winn, R. D. Meade, \emph{Photonic Crystals: Molding the Flow of Light}, (Princeton University Press, Princton, NJ, 2008).

\bibitem{polman15} A. F. Koenderink, A. Al\`{u}, and A. Polman, Science \textbf{348}, 516 (2015).
\bibitem{lodahl15} P. Lodahl, S. Mahmoodian, and S. Stobbe,  Rev. Mod. Phys. \textbf{87}, 347 (2015).

\bibitem{wu15} L.-H. Wu and X. Hu, Phys. Rev. Lett. \textbf{114}, 223901 (2015).
\bibitem{ma15} T. Ma, A. B. Khanikaev, S. H. Mousavi, and G. Shvets, Phys. Rev. Lett. \textbf{114}, 127401 (2015).
\bibitem{ma16} T. Ma and G. Shvets, New. J. Phys. \textbf{18}, 025012 (2016).
\bibitem{longphotonic} L. Xu, H. Wang, Y. D. Xu, H. Y. Chen, and J.-H. Jiang, arXiv:1601.03168.
\bibitem{dong15}X.-D. Chen and J.-W. Dong, arXiv:1602.03352.



\bibitem{neto09} A. H. Castro Neto, F. Guinea, N. M. R. Peres, K. S. Novoselov, and A. K. Geim, Rev. Mod. Phys. \textbf{81}, 109 (2009).

\bibitem{sakoda01} K. Sakoda, \emph{Optical Properties of Photonic Crystals}, (Springer-Verlag, Berlin, Germany, 2001).

\bibitem{palik97} E. D. Palik, \emph{Handbook of Optical Constants of Solids I-III}, (Academic Press, San Diego, CA, 1997).

\bibitem{painter99} O. Painter \emph{et al}., Science \textbf{284}, 1819 (1999).
\bibitem{englund05} D. Englund \emph{et al}., Phys. Rev. Lett. \textbf{95}, 013904 (2005).
\bibitem{kim13} H. Kim, R. Bose, T. C. Shen, G. S. Solomon, and E. Waks, Nature Photon. \textbf{7}, 373 (2013).

\bibitem{johnson99} S. G. Johnson, S. Fan, P. R. Villeneuve, J. D. Joannopoulos, and L. A. Kolodziejski, Phys. Rev. B \textbf{60}, 5751 (1999).


\bibitem{suppl} For more details, see appendix.

\bibitem{bernevig06} B. A. Bernevig, T. L. Hughes, and S.-C. Zhang, Science \textbf{314}, 1757 (2006).

\bibitem{petersen14} J. Petersen, J. Volz, A. Rauschenbeutel, Science \textbf{346}, 67 (2014).
\bibitem{sollner15} I. S\"{o}llner \emph{et al.}, Nature Nanotech. \textbf{10}, 775 (2015).





\bibitem{albert15} V. V. Albert, L. I. Glazman, and L. Jiang, Phys. Rev. Lett. {\bf 114}, 173902 (2015).

\bibitem{press08} D. Press, T. D. Ladd, B. Zhang, and Y. Yamamoto, Nature \textbf{456}, 218 (2008).
\bibitem{berezovsky08} J. Berezovsky, M. H. Mikkelsen, N. G. Stoltz, L. A. Coldren, and D. D. Awschalom, Science \textbf{320}, 349 (2008).
\bibitem{sun16} S. Sun, H. Kim, G. S. Solomon, and E. Waks,  	arXiv:1506.06036.


\bibitem{xia14}  F. Xia, H. Wang, D. Xiao, M. Dubey, and A. Ramasubramaniam, Nature Photon. \textbf{8}, 899 (2014).




\bibitem{aharonovich11} I Aharonovich \emph{et al}., Rep. Prog. Phys. \textbf{74}, 076501 (2011).
\bibitem{aharonovich11b} I. Aharonovich, A. D. Greentree, and S. Prawer, Nature Photon. \textbf{5}, 397 (2011).































\bibitem{haldane} D. N. Sheng, Z. Y. Weng, L. Sheng, and F. D. M. Haldane,	Phys. Rev. Lett. {\bf 97}, 036808 (2006).

\bibitem{ashcroft} N. W. Ashcroft and N. D. Mermin, \emph{Solid State Physics}, (Harcourt College Publishers, Fort Worth, 1976).

\bibitem{jackiw} R. Jackiw and C. Rebbi, Phys. Rev. D {\bf 13}, 3398 (1976).


\end{thebibliography}
\end{document}